\documentclass[prb,twocolumn,preprintnumbers,amsmath,amssymb]{revtex4}
\usepackage{graphicx,color, dsfont}
\usepackage{dcolumn}
\usepackage{bm}

\newcommand{\be}{\begin{equation}}
\newcommand{\ee}{\end{equation}}
\newcommand{\bea}{\begin{eqnarray}}
\newcommand{\eea}{\end{eqnarray}}

\newcommand{\bwt}{\begin{widetext}}
\newcommand{\ewt}{\end{widetext}}
\newcommand{\ham}{\mathcal{H}}

\newcommand{\ra}{\rangle}
\newcommand{\la}{\langle}
\newcommand{\bsb}{\begin{subarray}}
\newcommand{\esb}{\end{subarray}}

\newcommand{\tw}{\tilde{\omega}}

\begin{document}
\title{\bf Nano-engineered non-uniform strain in graphene}
\author{M. Neek-Amal$^{1,2}$, L. Covaci$^2$ and F. M. Peeters$^2$ }
\affiliation{$^1$Department of Physics, Shahid Rajaee University,
Lavizan, Tehran 16785-136, Iran.\\$^2$Departement Fysica,
Universiteit Antwerpen, Groenenborgerlaan 171, B-2020 Antwerpen,
Belgium.}
\date{\today}
\begin{abstract}
Recent experiments showed that non-uniform strain can be produced by
depositing graphene over pillars. We employed atomistic calculations
to study the non-uniform strain and the induced pseudo-magnetic
field up to 5000 Tesla in graphene on top of nano-pillars.  By
decreasing the distance between the nano-pillars a complex
distribution for the pseudo-magnetic field can be generated.
Furthermore, we performed tight-binding calculations of the local
density of states (LDOS) by using the relaxed graphene configuration
obtained from  the atomistic calculations. We find that the
quasiparticle LDOS are strongly modified near the pillars, both at
low energies showing sub-lattice polarization, and at high energies
showing shifts of the van Hove singularity. Our study shows that
changing the specific pattern of the nano-pillars allows us to
create a desired shape of the pseudo-magnetic field profile while
the LDOS maps provide an input for experimental verifications by
scanning tunneling microscopy.\end{abstract} \maketitle

 Graphene is a newly discovered atomic
thin two-dimensional honeycomb lattice consisting of carbon
atoms~\cite{1}. It is a zero gap semimetal with a conical band
structure where the conduction and valence bands touch each other at
the Dirac point~\cite{14}. Nano-engineered non-uniform strain
distribution in graphene is a promising road to generate a band gap
and a pseudo-magnetic field. 
 Scanning tunneling microscopy (STM) measurements have shown strain-induced Landau
levels~\cite{24} which correspond to a large pseudo-magnetic field.
Shear strain is essential and neither uniaxial nor isotropic strain
produces a strong uniform pseudo-magnetic field~\cite{23}.

Graphene's highly responses to external forces resulting in
mechanical deformations. Over the last few years there have been
many efforts to control graphene's electronic properties by
strain~\cite{32,33,34}. Elastic deformations create a
pseudo-magnetic field which acts on graphene's massless charge
carriers~\cite{naturegoonia,revmodphys,PRBmidgap}. The resulting
variation of the hopping energies can be viewed as an induced
pseudo-magnetic field which enters in the Dirac equation.
Engineering of the right topology of the induced pseudo-magnetic
field provides symmetrical magnetic confinement which confines
electrons in specific regions in space~\cite{PRBrapid}.

Recently, it was predicted that  non-uniform strain may lead to a
considerable energy gap and a large gauge field that effectively
acts as a uniform magnetic field ~\cite{35}. Recently, Tomori
\emph{et al} used pillars made of a dielectric material (electron
beam resist) which were placed on top of a substrate which is then
overlayed with graphene to generate non-uniform strain on a
micro-scale~\cite{38}. The graphene sections which are located
between the pillars are attached to the substrate and the size and
separation of the pillars control the strength and  distribution of
the strain.  The length scale in the experiment was micronmeteres
and SiO$_2$~was used as the substrate~\cite{38}.

Here we study non-uniform strain at the atomistic scale where the
continuum approach is no longer applicable. We also study the local
density of state (LDOS) maps using the relaxed graphene
configuration as input for tight-binding calculations. We find very
strong non-uniform pseudo-magnetic fields that can be created by
depositing graphene on a substrate decorated with nano-scale pillars
and find that the quasiparticle LDOS are strongly modified near the
pillars. The optimum configuration of graphene over such
nano-pillars depends on the imposed boundary conditions. The induced
pseudo-magnetic fields are larger than 1000 Tesla and are spatially
distributed around the nano-pillars. Decreasing the distance between
the nano-pillars alters the six-fold symmetry of the pseudo-magnetic
field distribution and results in a new configuration of magnetic
confinement for the charge carriers on the graphene around the
nano-pillars. Our study shows the LDOS maps around the pillars,
which can be experimentally verified by STM.


\textbf{\emph{Atomistic model}}. Classical atomistic molecular
dynamics simulation (MD) is employed to find the optimum
configuration of large flakes of graphene (GE) over the
nano-pillars. The second generation of Brenner's bond-order
potential is employed and is able to describe covalent sp$^3$ bond
breaking and the formation of associated changes in atomic
hybridization within a classical potential~\cite{brenner2002}. The
van der Waals (vdW) interaction between GE and
nano-pillars/substrate is modeled by employing the Lennard-Jones
(LJ) potential~\cite{MD2010,neek2010,PRB2010,ACSnano}.

\begin{figure}
\begin{center}
\includegraphics[width=\columnwidth]{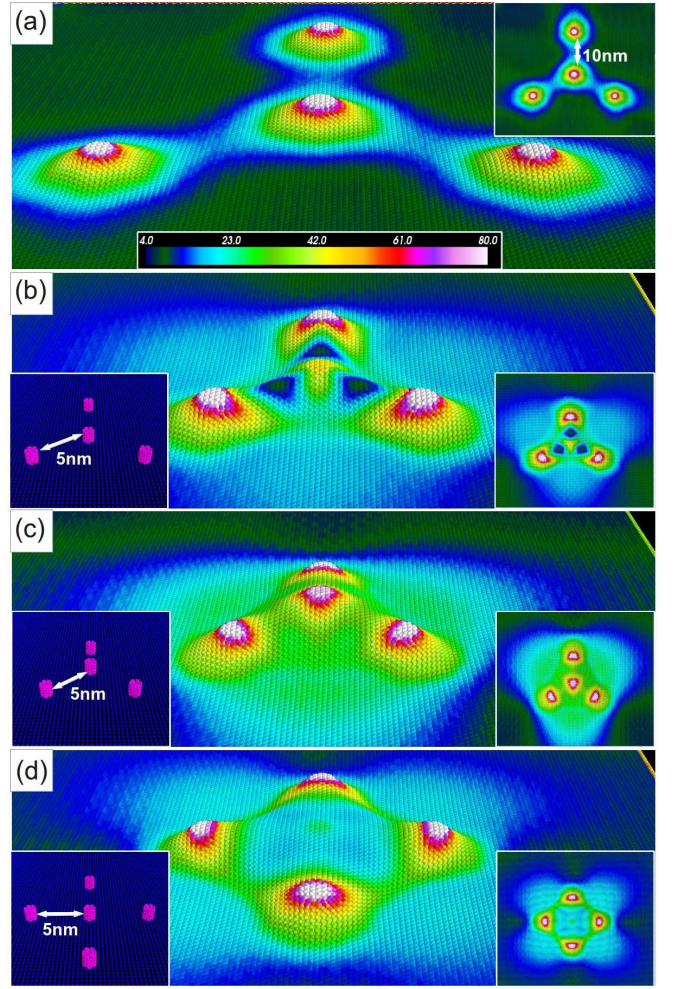}
\caption{(Color online)   The optimal configuration of graphene  on
top of double wall armchair carbon nanotubes (nano-pillars)
and a square lattice substrate (right-insets shows a top view).
The left-insets show the nano-pillars and the substrate. In (a) the distance between
pillars is 10\,nm while in (c-d) it is 5\,nm. All pillars are at the same height except (c) where the central pillar is 2nm higher. Colors indicate the
scaled stress distribution, i.e. white represents highest stress and dark-green lowest
stress.}
\label{fig1}
\end{center}
\end{figure}

In order to model the substrate, a (100) surface with lattice
parameter equal to $\ell$=3\,\AA ~is assumed with LJ parameters
$\sigma_S$ and $\epsilon_S$. The density of the sites in the
substrate is
$\Sigma_S=\ell^{-2}$ and the number of atoms is 13700. 
Nano-pillars are double-wall armchair carbon nanotubes (DWCNT) taken
with (3,3) and (6,6) indexes including 144 atoms (see left insets
in~Fig~\ref{fig1}). The number of atoms in the graphene sheet
 is 44800 which is equivalent to a sheet of size
34.8$\times$34.43\,nm$^2$. We assume that both the substrate and
nano-pillar atoms are rigid during the simulation.

To model the interaction between two different types of atoms such
as carbon atom (C) and substrate atom (S), we adjust the LJ
parameters using the equations $\epsilon_T\,=\, \sqrt{\epsilon_C
\epsilon_S}$ and $\sigma_T\,=(\sigma_{C}+\sigma_S)/2$. The
parameters for carbon are
$\sigma_C\,=\,3.369\,$\AA~and~$\epsilon_C\,=\,2.63$\,meV. For the
substrate atoms we took $\sigma_S=$3.5\AA~ and $\epsilon_S$
=10.0\,meV which are typical for insulators, e.g.
SiO$_2$~\cite{MD2010}.

The atomic stress experienced by each $i^{th}$ atom can be expressed
as \cite{PRB2004,carbon}
\begin{equation}
\eta^{i}_{\mu \nu}=\frac{1}{\Omega}\left(\frac{1}{2}m v^{i}_{\mu}
v^{i}_{\nu}+\sum_{j \neq i} r^{\nu}_{ij} F^{\mu}_{ij}\right),
\end{equation}
where the inner summation is over all the carbon atoms which are
neighbors of the $i^{th}$ atom which occupies  a volume $\Omega=
4\pi a_0^3/3$. The quantities $m$ and $v^i$ denote the mass and
velocity of $i^{th}$ atom  and the scaler $r^{\nu}_{ij}$ is the
$\nu$ component of the distance between atoms `i' and `j'.
$F^{\mu}_{ij}$ is the force on $i^{th}$ atom due to atom $j^{th}$ in
the $\mu$ direction. We used this expression to calculate the stress
on each atom. In order to be able to visualize the stress
distribution on the GE atoms, we colored the atoms using a
dimensionless invariant quantity~\cite{rafii}
$J_2=\frac{1}{6}[(\eta_{xx}-\eta_{yy})^2+(\eta_{yy}-\eta_{zz})^2+(\eta_{zz}-\eta_{xx})^2+6(\eta_{xy}^2+\eta_{xz}^2+\eta_{yz}^2)]$,
i.e. dark green (white) is related to a minimum (maximum) value of
$J_2$.

\textbf{\emph{Strain induced pseudo-magnetic field}}. Coupling the
Dirac equation, which governs the low energy electronics of graphene,
to the curved surface is a common way to study the effects of
graphene's curved geometries on its corresponding electronic
properties~\cite{revmodphys,PRBmidgap}. The metric of the curved
surface describes the curvature of the surface. The origin of the
deformations are external stresses that deform graphene so that the
nearest neighbor distances become non-equal. The latter results in
modified hopping parameters, which are now a function of the atomic
positions $t(\textbf{r})$~\cite{PRBrapid}. Assuming small atomic
displacements ($\textbf{u}=\textbf{r}'_i-\textbf{r}_i<a_0$) and
rewriting the Dirac Hamiltonian in the effective mass approximation
with non-equal hopping parameters yields the induced gauge fields:
\begin{equation}\label{gauge}
\textbf{A}=\frac{2\beta\hbar}{3a_0 e}(u_{xx}-u_{yy},-2u_{xy}),
\end{equation}
where $\beta\sim 3$ is a constant and $u_{\alpha\beta}$ is the
strain tensor including out of plane
displacements~\cite{revmodphys}. The corresponding pseudo-magnetic
field perpendicular to the $x-y$ plane is obtained as
\begin{equation}\label{Magnetic}
B=\partial_y A_x-\partial_x A_y.
\end{equation}
This is the  pseudo-magnetic field that an electron experiences in
the K-valley. We will find $B$ by making the necessary
differentiations numerically in the case of supported boundary
conditions. Here we found that the major contribution is due to the
out-of-plane terms, which appear mainly around the deformed parts.
The other in-plane terms contribute less to the pseudo-magnetic
field around the deformed parts particularly when the system is
large as compared to the deformed regions.

\textbf{\emph{Tight binding model}} .The electronic properties are described by a
 tight-binding Hamiltonian for the $\pi$ carbon orbitals.
The minimal Hamiltonian, which describes the low-energy band
structure is: \be \label{eq:hamil} \ham=\sum_{\la i,j \ra, \sigma}
-t(r_{ij}) c_{i\sigma}^\dagger c_{j\sigma} + h.c. \ee where
$c_{i\sigma}^\dagger$ ($c_{j\sigma}$) creates (destroys) an electron
at site $i$ ($j$). The sum  runs over nearest neighbors pertaining
to opposite sub-lattices $\la i,j \ra$ and the
 electron spin, $\sigma$. In the following we will ignore the spin degrees of freedom since no
 spin-flipping term is present in the Hamiltonian.

 \begin{figure*}[ttt]
 \includegraphics[width=\textwidth]{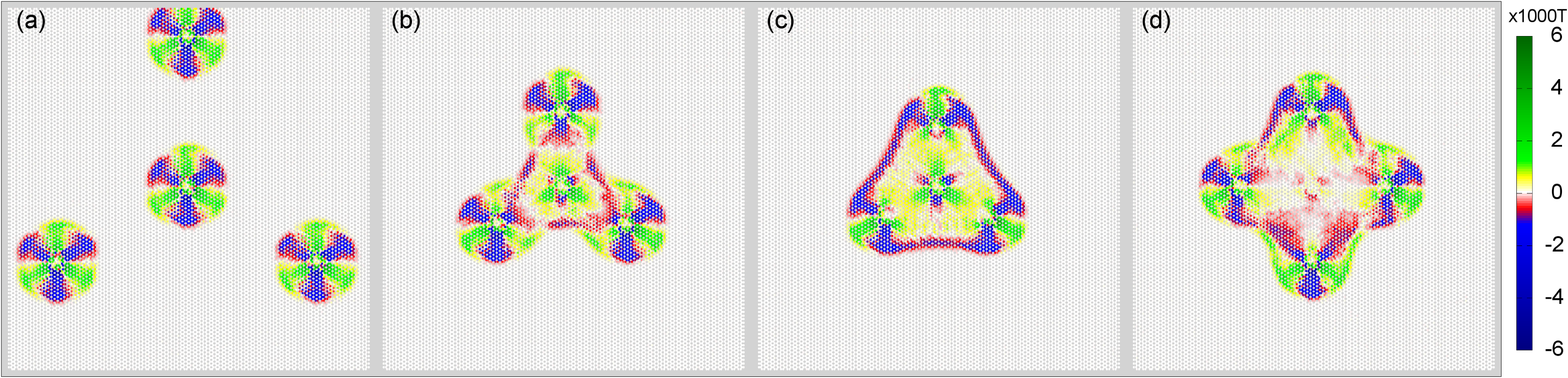}\\
 \includegraphics[width=\textwidth]{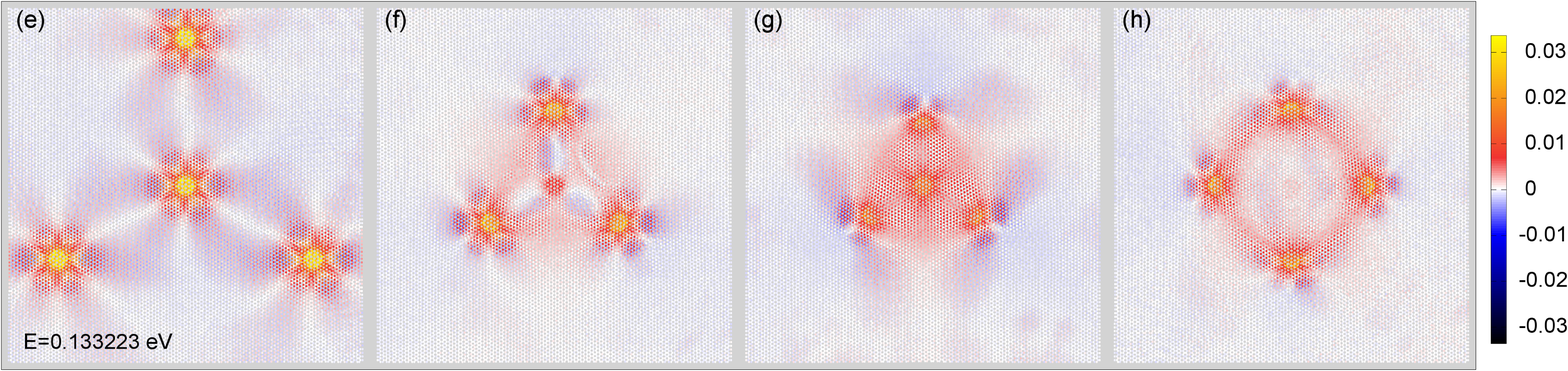}\\
  \includegraphics[width=\textwidth]{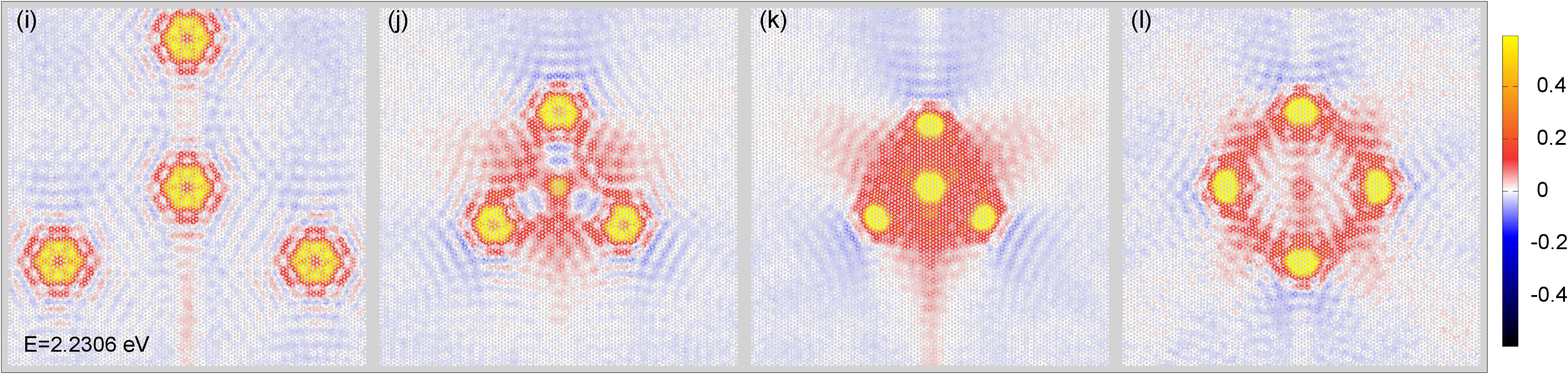}
\caption{(Color Online). (a-d) Pseudo-magnetic fields in one Dirac
cone for 4 different pillar configurations shown in Fig.~1,
respectively. (e-h) Low energy (E=0.1332eV) LDOS map and (i-l)
high energy (E=2.2306eV) LDOS map for the same 4 configurations.
Note that the bulk unstrained LDOS is subtracted from the LDOS maps.
} \label{fig2}
 \end{figure*}
The strain is included in the modified hopping amplitudes between
$\pi$ orbitals, $t_\pi(r_{ij})$, according to the empirical relation
$t_\pi(r_{ij})=\gamma_0 \exp^{3.37(\frac{r_{ij}}{a_0}-1)}$, where
$\gamma_0=2.7eV$ and $a_0=1.42$\AA~is the equilibrium inter-carbon
distance~\cite{32}. This also gives a good approximation for the
next-nearest neighbor hopping amplitude. We also consider the effect
of misalignment of the $\pi$ orbitals due to the finite curvature.
This effect translates into the mixing of the $\pi$ and $\sigma$
orbitals. Depending on the local curvature, the modified hopping
amplitude is: \be
t(r_{ij})=t_{\pi}\sin(\theta_i)\sin(\theta_j)cos(\phi)-t_{\sigma}\cos(\theta_i)\cos(\theta_j),
\ee where $\theta_i$ and $\theta_j$ are the angles formed by the
normals at each atom, $\mathbf{n}_i$ and $\mathbf{n}_j$, with the
inter-atomic distance $\mathbf{r}_{ij}$ and $\phi$ describes the
angle formed by the normal $\mathbf{n}_j$ and the plane defined by
$\mathbf{n}_i$ and $\mathbf{r}_{ij}$ \cite{zozulenko, seba, loss}.
The strain configuration and the curvature are extracted from the
relaxed position of the graphene sheet obtained from our molecular
dynamics simulation. For the systems considered here the effect of
curvature is small when compared to the effect of strain on the
hopping amplitudes.

Since the strain is inhomogeneous it is not possible to use any
symmetries in the calculations of the electronic properties.
Therefore, the system size considered here (44800 carbon atoms)
becomes prohibitively large for an exact digitalization of the
Hamiltonian. Instead we will numerically obtain an approximation of
the Green's function by using a Chebyshev expansion within the
Kernel Polynomial method \cite{weise2006, fehske, covaci2010,
covaci2012}. The Green's function is defined as: \be
\label{eq:green1} G_{ij}(\omega)=\la c_{i} | \hat{G}(\omega) |
c_{j}^\dagger  \ra \ee where $\hat{G}(\omega+i \eta)=[\omega +
i\eta-\ham]^{-1}$.

First a scaling of the excitation energies is performed, e.g.
 $\tilde{\ham}=(\ham-\mathds{1}b)/a$, $\tilde{\omega}=(\omega-b)/a$
 where $a=(E_{max}-E_{min})/(2-\eta)$ and $b=(E_{max}+E_{min})/2$,
  where $\eta>0$ is a small number. Following Refs.~[\onlinecite{covaci2010,weise2006}],
  the Green's function's components can be expressed as an expansion written in terms of Chebyshev polynomials:
\be
\label{eq:gf1}
G_{ij}(\tw)=\frac{-2 i}{\sqrt{1-\tw^2}} \sum_{n=0}^\infty a_n(i,j) e^{-i\: n \cdot \arccos(\tw)}
\ee
where the coefficients $a_n(i,j)= \la c_i| v_n \ra$ can be obtained by an iterative
 procedure involving repeated applications of the Hamiltonian on iterative vectors $| v_n \ra$:
\be
 |v_{n+1} \ra = 2\ham |v_{n} \ra - | v_{n-1} \ra,
\ee
where $|v_0 \ra = |c_j^\dagger \ra$ and $|v_{-1} \ra = 0$.  Significant computational
 speed-up is achieved when the computations are done on graphical processing units (GPU), i.e.
  video cards. The computations are performed on Nvidia GeForce GTX 580 cards.

The physical properties that can be straightforwardly extracted from
the Green's functions are the local density of states (LDOS): \be
N_i(\omega)=-\frac{2}{\pi} Im[G_{ii}(\omega)], \ee the factor of $2$
appears due to the summation over the spin components.

\textbf{\emph{Results and discussion}}. At the start of our
simulation we put graphene on top of the nano-pillars at
$h_0=1.4$\,nm. The substrate is at zero height and the nano-pillars
are located between graphene and the substrate. We have investigated
two particular patterns of nano-pillars: i) five DWCNTs which have
in-plane coordinates (0,0) and ($\pm d$,$\pm d$) with $d=5,10$\,nm,
and ii) four DWCNTs at (0,0), ($\pm d \sqrt{3}/2$,$\pm d/2$) and
(0,-$d$). The height of DWCNTs was set to be 1\,nm (except in
Fig.~\ref{fig1}(c) which is 1.5\,nm). In order to prevent crumpling
at the boundaries we only allow the boundary atoms to vibrate in the
$z$-direction.

In Fig.~\ref{fig1} we show the optimal configuration of GE on top of
four (a-c) and five (d) DWCNTs. Right insets in
Figs.~\ref{fig1}(a-d) show a top view. Notice that the stress
distribution is mainly concentrated around the nano-pillars as
expected.
 For the configuration presented in  Fig.~\ref{fig1}(a) the pillars are 10nm apart.
 Due to the vdW interaction the graphene sheet will stick to the substrate except around the
 pillars where the shape is close to a Gaussian even though deviations from an isotropic description exists.
  In the atomic limit, a slight anisotropy appears, the graphene sheet bends mostly in the zig-zag direction
   making the shape of the deformation hexagonal. In Figs.~\ref{fig1}(b-d) the pillars are closer
    together (i.e. 5nm). Due to its large bending rigidity, the graphene sheet will be suspended over the substrate
     in the regions between the pillars. Depending on the pillar configuration, various stain configurations
     are achieved. If all the pillars have the same height, Figs.~\ref{fig1}(b) and 1(d), most of the stress is
      obtained at the pillar location and where graphene sticks to the substrate. If one of the pillars is higher,
      Figs.~\ref{fig1}(c), besides the maximal stress at the pillar location, high stress is also obtained
      throughout the suspended sheet.

The corresponding pseudo magnetic field profiles generated by the
strain configurations are shown in Figs.~\ref{fig2}(a-d). When the
deformations are isolated, Fig.~\ref{fig2}(a), the gauge field and
the pseudo magnetic field exhibit  six fold
symmetry~\cite{PRBrapid,23} similar to a Gaussian deformation,
$h(x,y)=G\exp(-\frac{x^2+y^2}{2\sigma^2})$. The continuum theory
predicts that the pseudo-magnetic gauge field is
$\textbf{A}=\frac{h(x,y)^2}{2\sigma^4}(x^2-y^2,-xy)$ and the
pseudo-magnetic field is $B=\frac{h(x,y)^2
}{\sigma^6}(x^2+y^2)\sin(3\theta)$, where $\theta$ is the azimuthal
angle. Large pseudo-magnetic fields on the order of thousands of
Teslas are obtained. When the graphene sheet is suspended,
Figs.~\ref{fig2}(b-d), the six-fold symmetry survives near the
pillars but more complex pseudo-magnetic field profiles can be
obtained, from large fields throughout the suspended sheet,
Fig.~\ref{fig2}(c), to fields localized only near the edges of the
suspended sheet,  Fig.~\ref{fig2}(d). As seen from
Figs.~\ref{fig2}(c,d) the closer the pillars, triangular and
rectangular magnetic field profile is created within the position of
the pillars. In Figs.~\ref{fig2}(c) there is a high magnetic field
region at the center and  the electron  can not pass through this
region.

In order to investigate the effect of the strain on the electronic properties,
 we input the relaxed positions of the atoms obtained from the atomistic simulation into
 the tight-binding model in order to find the LDOS maps around the pillars. These
  are shown in Figs.~\ref{fig2}(e-h) for $E=0.1332$eV and in Figs.~\ref{fig2}(i-l)
   for $E=2.2306$eV. Two regimes can be observed, depending on the energy. For low energies,
    the pseudo-magnetic field will induce sub-lattice polarized states localized either near
    the pillars, Fig.~\ref{fig2}(e), or in the regions with large pseudo-magnetic fields, Fig.~\ref{fig2}(g).
     In the five-pillar configuration, Fig.~\ref{fig2}(h), these low energy states are mostly localized near
     the edges of the suspended region. Interference patterns which depend on the energy are observed \cite{supp}.
     A very different effect, which is not described by the low energy
      Dirac approximation \cite{23}, is related to the shift of the van Hove singularity seen in unstrained
      graphene at $E=\gamma_0=2.7$eV.
       Because of strain, the hopping parameters will be modified, therefore locally shifting the van Hove
        singularity. This is observed in Figs.~\ref{fig2}(i-l), where the enhancement of the LDOS is correlated
         with the stress and is enhanced where the stress is larger. Additional interference patterns appear
         between the pillars. The deviation from the Gaussian shape of the LDOS modification, i.e. the hexagonal shape, of the isolated
         deformations is also obvious from Fig.~\ref{fig2}(i).

\textbf{\emph{Conclusions}}. By combining molecular dynamics
simulations with tight binding calculations we have shown how strain
can be manipulated at the nano-scale. Isolated pillars show a
six-fold symmetric pseudo-magnetic field and LDOS modification. By
decreasing the distance between the pillars, the six fold symmetry
of the pseudo-magnetic field is altered and a complex field profile
appears within the suspended area. We found that by modifying the
inter-pillar distances and the pillar heights one can design a
particular desired magnetic field profile. Modifications of the
hopping parameters due to changes in the C-C distances induced by
strain, modify the LDOS around the deformations of the graphene
sheet. Verifications of the six-fold symmetry of the LDOS near
pillars could be easily confirmed with STM experiments.

\emph{{\textbf{Acknowledgment}}}.   This work was supported by the
Flemish Science Foundation (FWO-Vl) and  the EuroGRAPHENE project
CONGRAN.

\end{document}